# Radiance enhancement and shortwave upwelling radiative flux methods for efficient detection of cloud scenes

Rehan Siddiqui [1,2,3,4], Rajinder K. Jagpal [1,3,4], Sanjar M. Abrarov [2,3,4] and Brendan M. Quine [1,2]

[1] Dept. Physics and Astronomy, York University, 4700 Keele Street, Toronto, M3J 1P3, Canada

[2] Dept. Earth and Space Science and Engineering, York University, Toronto, M3J 1P3, Canada

[3] Epic College of Technology, 5670 McAdam Rd., Mississauga, L4Z 1T2, Canada

[4] Epic Climate Green (ECG) Inc., 23 Westmore Dr., Unit 310, Toronto, M9V 3Y7, Canada

**ABSTRACT**

The description, imagery and interpretation of cloud scenes by remote sensing datasets from Earth-orbiting satellites have become a great debate for several decades. Presently, there are many models for cloud detection and its classifications have been reported. However, none of the existing models can efficiently detect the clouds within the small band of shortwave upwelling radiative wavelength flux (SWupRF) in the spectral range from 1100 to 1700 nm. Therefore, in order to detect the clouds more effectively, a method known as the radiance enhancement (RE) can be implemented (Siddiqui et al., 2015). This article proposes new approaches how with RE and SWupRF methods to distinguish cloud scenes by space orbiting Argus 1000 micro-spectrometer utilizing the GENSPECT line-by-line radiative transfer model (Quine and Drummond, 2002; Siddiqui, 2017). This RE approach can also be used within the selected wavelength band for the detection of combustion-originated aerosols due to seasonal forest fires.

## 1. Introduction

The World Climate Research Program (WCRP) was the first project established under the International Satellite Cloud Climatology Project (ISCCP) in 1982 to collect and analyse a globally uniform satellite radiance dataset to produce new cloud climatology (Schiffer and Rossow, 1983). Clouds are central occurrences that provide a link between the two key energy exchange processes that determine the Earth climate, particularly the solar – terrestrial and solar – water radiance exchanges (Rossow and Garder, 1993). Cloud detection from remote sensing dataset continues to be one of the most perspective research areas since 80's (see for example Ebert, 1987; Li and Leighton, 1991; Cutillo et al., 2004; Song et al., 2004; Li and Li, 2008; Ghosh et al., 2012; de Leeuw et al., 2012; Tang et al., 2013; Illingworth et al., 2015; Guo et al., 2015; Siddiqui et al., 2015, 2017; Siddiqui, 2017; Xi et al., 2015; Someya et al., 2016; Schreier and Suselj, 2016). Clouds are generally characterised by higher reflectance and lower temperature than the underlying Earth surface (Ackerman et al., 1998). Clouds with radiation energy always play most important role through absorption and scattering of photonic radiance within different atmospheric layers due to water vapour, carbon dioxide, oxygen and aerosols. Knowledge and monitoring of the Earth radiation budget is essential for improving our understanding of the Earth climate and potential climate change (Hatzianastassiou et al., 2005). The Earth surface net shortwave radiation, – the difference between the incoming and outgoing SW radiations, – represents the amount of solar radiation absorbed by surface (Inamdar and Guillevic, 2015). Clouds are the main factor in modulating the Earth energy budget and the climate (Mitchell and Finnegan, 2009). Furthermore, clouds are highly important in evolution and dynamics of the Earth climate as well as the environmental impact on the Earth by heat transfer process (Siddiqui, 2017). In general, clouds take their ambient environment temperature that under normal lapse rate conditions usually decreases with increasing height in the atmosphere. Therefore, when emission is converted to equivalent black body temperature, it can be used to distinguish the presence of opaque clouds from warm surface with various threshold techniques (see e.g., Fournier et al., 2006). According to Jagpal et al. (2010) clouds and the other airborne particles absorbs or scatter a significant portion of the sunlight back to space (due to reflectance) before it transmits downwards the full atmospheric column, precluding full column of $CO_2$ measurements in the region occupied by opaque clouds.

The incoming solar radiation is attenuated as it penetrates into the atmosphere, reflects from the surface of the clouds and Earth, and travels back to space. In real atmosphere, the attenuation includes the molecular (Rayleigh) scattering, absorption by $CO_2$, $CH_4$, $CO$ and water vapours and droplets in a form of clouds (Siddiqui et al., 2016a). The most abundant greenhouse gas water vapour ($H_2O$) is particularly significant for clouds analysis. The attenuation also includes extinction (absorption as well as scattering) by aerosols and transmissive cirrus clouds, and partial reflection at components claimed by (Mao and Kawa, 2004). Clouds affect the path of photons through the atmosphere and, therefore, change optical depth within absorption band.

In cloud retrievals from a satellite, it is necessary to have a good estimate of the surface albedo. The reason is that the cloud detection is usually performed by comparing the measured reflectance with expected reflectance from the cloud scene (Li and Li, 2008). Cloud detection is a preliminary important step in most algorithms for processing radiance data that has been measured from satellites. In general, different clouds models are introduced in radiance transfer models and their influence on the radiance emitted from the Earth surface is estimated with respect to clear sky conditions at spectral regions as described by Cutillo et al. (2004).

One of the most efficient ways to represent accurately atmospheric variation with height is to divide the atmosphere into a large number of relatively thin homogeneous layers or cells where the required parametric values, assigned to each property of interest in each layer, are equal to the corresponding parametric values in the real atmosphere at each mid-point height of the specific layer (Quine and Drummond, 2002). The satellite instruments measures the radiances emitted or absorbed by the surface, atmosphere or clouds into the instruments line of sight captured by a small-size remote sensors. Radiance reflected by the top cloud layer depends on cloud thickness, cloud particle number density, size and shape (Mishchenko et al., 1996).

The near infrared (NIR) satellite instrument is one of the most efficient sources of detecting cloud scenes. The NIR cloud phase detection method developed by using MODIS algorithm (Pagano and Durham, 1993). Along with the launch of multi spectral satellite, e.g., TERRA, AQUA and MODIS (Xiong et al., 2009, 2010), multi-spectral synthesis methods are applied in cloud detection within NIR spectral region to detect cloud (Li and Li., 2008). Space orbiting Argus 1000 (Jagpal et al., 2010; Jagpal 2011; Siddiqui et al., 2015; Chesser et al., 2012) covering the wavelength band of 1100 nm to 1700 nm that falls under the category of NIR shortwave radiation band, can also be adopted to detect efficiently the cloud scenes.

The work presented in this paper mainly covers the retrieval of Argus 1000 dataset for the calculation of RE and total SWupRF (Siddiqui et al., 2017; Siddiqui, 2017) with help of the GENSPECT line by line radiative transfer model based on the different input parameters, such as cloud albedo, solar zenith angle (SZA), water vapour concentration, atmospheric concentration mixing ratios of $O_2$, $CO_2$ and $CH_4$ that significantly facilitate the procedure for efficient detection of the cloud scenes (Siddiqui, 2017).

## 2. Instrument profile and radiative transfer model

### 2.1. Observational pattern of Argus 1000 – a micro spectrometer

The Argus 1000 micro-spectrometer is shown in Figure 1. It was developed at York University, Canada in association with Thoth Technology Inc., is a part of the CanX-2 satellite's payload (Rankin et al. 2005; Sarda et al., 2006) and launched into space in 2008.

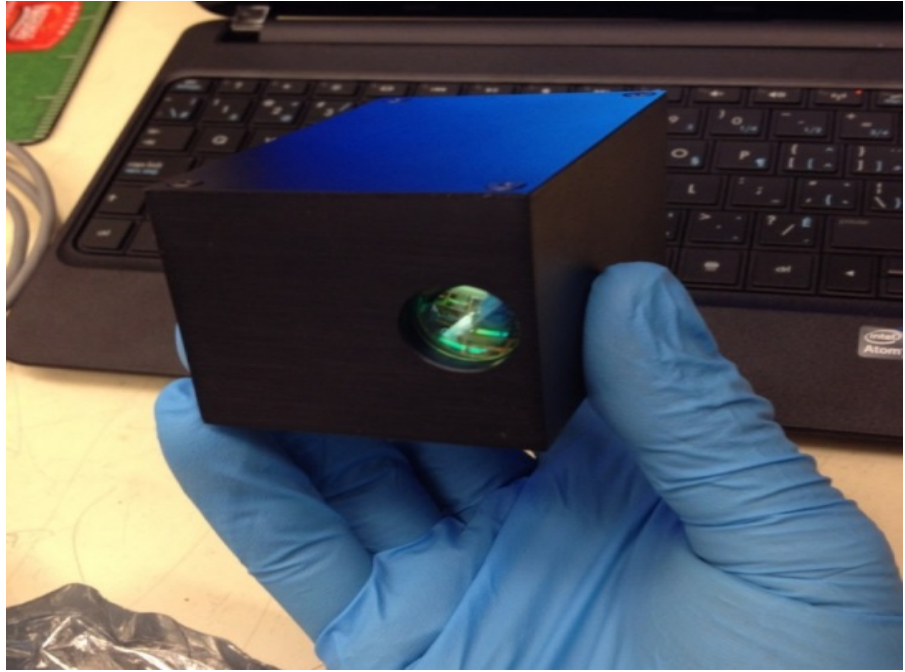

**Figure 1**. Argus 1000 spectrometer at the Space Engineering Laboratory, York University.

CanX-2 orbits in a low Earth orbit (LEO), 640 km above the Earth surface where Argus field of view (FOV) provides a spatial resolution of 1.4 km.

Figure 2 shows the both front and backend of CanX-2 configuration in real time view during the assembly process. The Argus 1000 micro-spectrometer operates in the near infrared (NIR) covering overall region from 900 nm to 1,700 nm with spectral resolution of about 4–6 nm (Jagpal et al., 2010). The Argus instrument provides a means to make measurements of upwelling radiation reflected to space by the Earth surface and atmosphere. Reflection spectra of sunlight from the Earth surface contain significant absorption features associated with the molecular absorption by particular gas species that can be used to predict the composition of the atmosphere (Quine and Drummond, 2002; Jagpal, 2011).

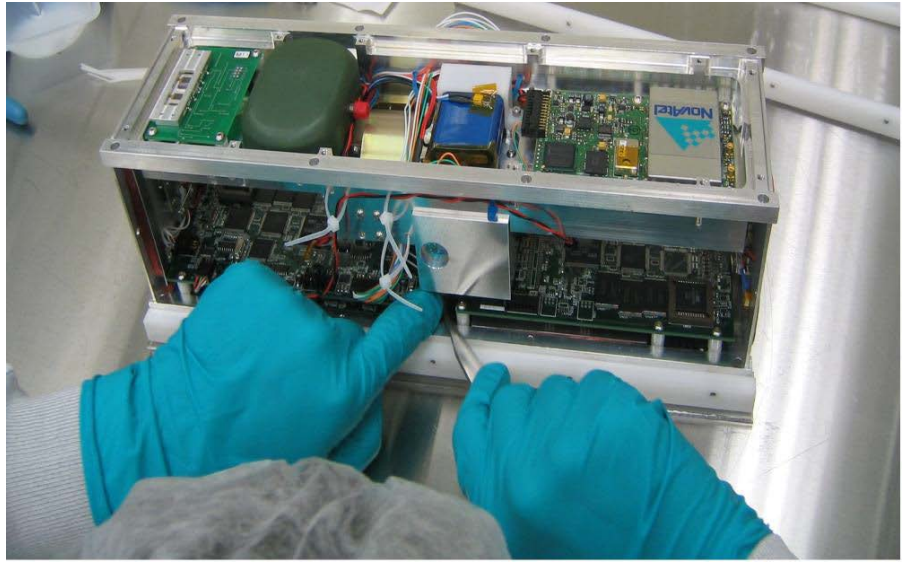

**Figure 2.** The front end view of engineering model Argus 1000 (pointed by arrow) attached with CanX-2 nano-satellite [Source UTIAS].

Argus 1000 records the NIR signature of the surface-troposphere amounts of the significant greenhouse gases oxygen $O_2$, carbon dioxide $CO_2$ and water vapour $H_2O$ in order to monitor anthropogenic pollution and to identify their sources including particulate matter $PM_{2.5}$ (Christopher and Gupta, 2010) in the atmosphere (Siddiqui et al., 2015; Quine and Drummond, 2002). Methane ($CH_4$), nitrous oxide ($N_2O$), carbon monoxide (CO) and hydrogen fluoride (HF) species also have absorption features in this spectral region of 900 nm to 1,700 nm (Jagpal, 2011). The instrument operates from space to record IR spectra of reflected solar radiation using a linear photodiode array that records the incident radiant energy (Jagpal, 2011). The measured spectra can be compared with IR absorption signatures obtained by linear path forward modelling of the atmospheric absorption process for determination of the various concentrations of absorbing species. In the absence of saturation or scattering effects, the amount of absorption depends upon the density of the absorber gas along the path. Therefore, the primary measurement objective of the instrument is to observe any changes in optical depth, associated with the variation of the following atmospheric gas species in the spectral interval 900–1,700 nm (11,111–5,882 cm$^{-1}$). Table 1 shows the observed absorption in this spectral range and their typical absorption strengths (Jagpal, 2011; Siddiqui, 2017).

**Table 1:** Species observed by Argus

| Observed target gas | Absorption wavelength (nm) | Comments |
|---|---|---|
| Oxygen ($O_2$) | 1260 | Very strong absorption due to $O_2$ abundance |
| Water ($H_2O$) | 900 1200 1400 | Dominant IR absorber 900-1700 nm |
| Carbon Dioxide ($CO_2$) | 1240 1420 1570 1600 | 1600 nm features are well isolated |
| Methane ($CH_4$) | 1660 | Low abundance in this spectral band |

The instrument was designed to take nadir observations of reflected sunlight from Earth surface and atmosphere (Jagpal, 2011). The nadir viewing geometry of Argus 1000 is of particular utility as this observation mode provides a useable probing data in regions that are partially cloudy or have surface topography representing practical interest. Argus team at the Space Engineering Laboratory at York University prepares the observation tables for the desired targets around the globe using the systems tool kits (STK) software. The Argus 1000 target list contains 35 global areas around the Earth. During many years of Argus 1000 space heritage, we have made over 300 reported observations over a series of land and ocean targets (Siddiqui et al., 2015), few examples of the selected Argus dataset for the detection of cloud scene are shown in Table 2 (Siddiqui, 2017).

Figures 3, 4 and 5 illustrate the Argus spectral profile of selected week per pass per observation numbers. All these spectra show a high or low radiance in contrast with each other within their selected NIR wavelength bands of interest of $O_2$, $H_2O$, $CO_2$ and $CH_4$.

**Table 2.** Argus selected week per pass per observations with Geo-location

| Week No., pass No. | Date | Selected observation numbers | Observations number with satellite Sun angle, Nadir angle, Lat. & Long. | Location |
|---|---|---|---|---|
| Week08_Pass61 | 2009Oct30 | 64,116,196,238 | OBS64:<br>Sat. nadir angle = 5.7579<br>Sat. sun angle = 35.3047<br>Lat. = 6.5100, Long. = 60.9622<br><br>OBS116:<br>Sat. nadir angle = 6.8112<br>Sat. sun angle = 32.3569<br>Lat. = 3.2055, Long. = 60.1266<br><br>OBS196:<br>Sat. nadir angle = 3.4581<br>Sat. sun angle = 29.2728<br>Lat. = -1.5265, Long. = 59.2479<br><br>OBS238:<br>Sat. nadir angle = 3.6066<br>Sat. sun angle = 29.1451<br>Lat. = -4.2681, Long. = 58.7476 | Arabian Sea and Seychelles |
| Week14_Pass52 | 2010Mar04 | 22,100,125 | OBS22:<br>Sat. nadir angle = 24.7152<br>Sat. sun angle = 58.1136<br>Lat. = 47.3740, Long. = -77.7286<br><br>OBS100:<br>Sat. nadir angle = 23.8760<br>Sat. sun angle = 54.9109<br>Lat. = 42.8413, Long. = -79.7189<br><br>OBS125:<br>Sat. nadir angle = 21.8777<br>Sat. sun angle = 54.5434<br>Lat. = 42.2893, Long. = -79.9453 | Toronto/ Kitcisakik (Canada) |
| Week75_Pass43 | 2013Aug14 | 30,43,65 | OBS30:<br>Sat. nadir angle  =  1.7477<br>Sat. sun angle      = 38.2453<br>Lat. = 28.9233, Long. = 147.5652<br><br>OBS43:<br>Sat. nadir angle = 1.6689<br>Sat. sun angle = 37.9789<br>Lat. = 25.5822, Long. = 146.8083<br><br>OBS65:<br>Sat. nadir angle = 1.6877<br>Sat. sun angle = 38.1652<br>Lat. = 19.9246, Long. = 145.5481 | North Pacific Ocean |

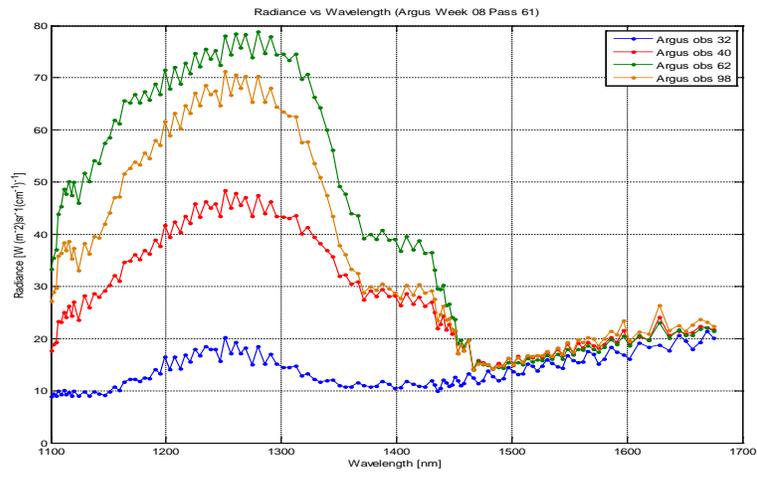

**Figure 3.** Argus spectra – radiance vs. wavelength of week 08 pass 61 with selected observation numbers.

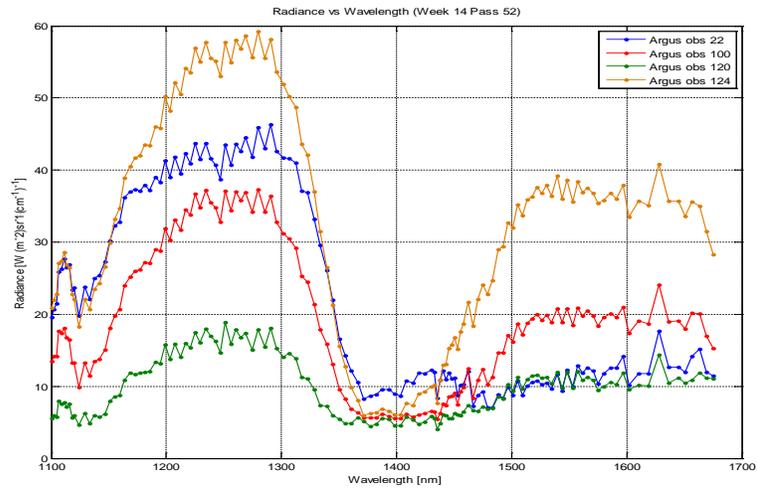

**Figure 4.** Argus spectra – radiance vs wavelength of week 14 pass 52 with selected observation numbers.

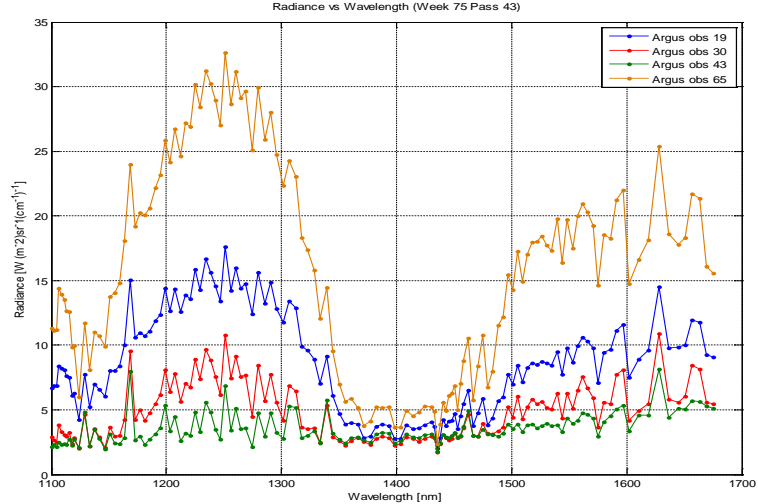

**Figure 5.** Argus spectra - radiance vs wavelength of week 75 pass 43 with selected observation numbers.

## 2.2. GENSPECT – a line by line radiative transfer model

The GENSPECT is a line-by-line radiative transfer algorithm for absorption, emission, and transmission for a wide range of atmospheric gases. GENSPECT uses a variable frequency grid to compute absorption parameters to a specified accuracy by using the Voigt profile (Abramowitz and Stegun 1972; Abrarov et al., 2010a, 2010b). The HITRAN line strengths (Duggan et al., 1993) are pre-adjusted for normal isotopic abundances and tabulated, to model an Earth atmosphere with natural abundance. Given information including gas types and amounts, pressure, path length, temperature, and frequency range for an atmosphere, the GENSPECT model computes the spectral characteristics of the gas. GENSPECT employs a unique computation algorithm that maintains a specified accuracy for the calculation by pre-computing where a line function may be interpolated without a reduction in accuracy (Quine and Drummond, 2002). The approach employs a binary division of the spectral range, and calculations are performed on a cascaded series of wavelength grids, each with approximately twice the spectral resolution of the previous one. The GENSPECT error tolerances are 0.01%, 0.1%, and 1%, which may be selected according to the application (Quine and Drummond, 2002). GENSPECT has been used previously to compute synthetic spectra for data retrieval; collected by Earth observing instruments deployed in the air, in space and on the ground including MOPITT-A, MOPITT, ACE-FTS, and MAESTRO (McKernan et al., 2002; Quine and Drummond, 2002). In order to corroborate satellite observational results, radiative transfer simulations are also performed by using GENSPECT model. The general input for the radiative transfer is the radiance enhancement (RE) and upwelling radiative flux in comparisons with Argus observational dataset taken out as different profile of temperatures, solar zenith angle, mixing ratio concentrations of $O_2$, $H_2O$, $CO_2$, $CH_4$ and surface albedo.

Figures 6 and 7 show few spectral profiles of GENSPECT-synthetic model with albedo 0.3 and 0.9 at different water level concentrations. Both spectra are in a reasonable agreement with absorption features of $O_2$, $H_2O$, $CO_2$ and $CH_4$ within their selected NIR wavelength bands of interest. These two figures also show the dominant increase of radiance shift by changing water vapour concentration, surface albedo and altitudes from surface to reflecting medium.

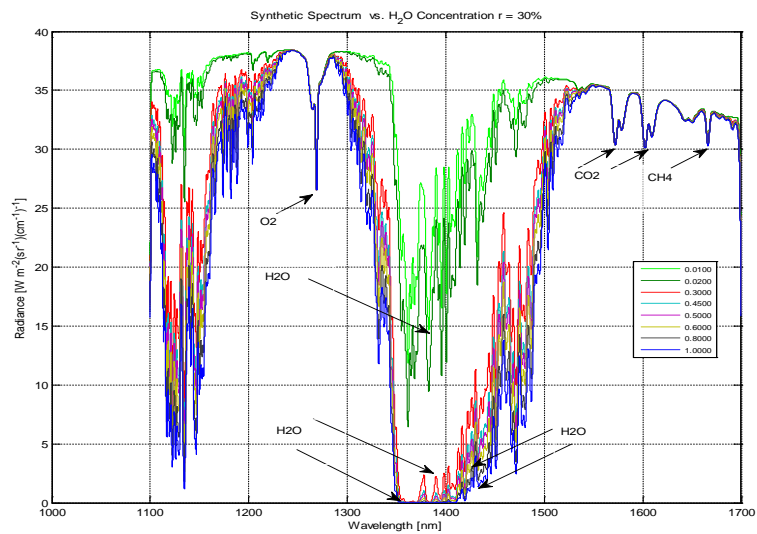

**Figure 6.** GENSPECT – synthetic spectrums with $H_2O$ from 1% to 100% and albedo = 0.3.

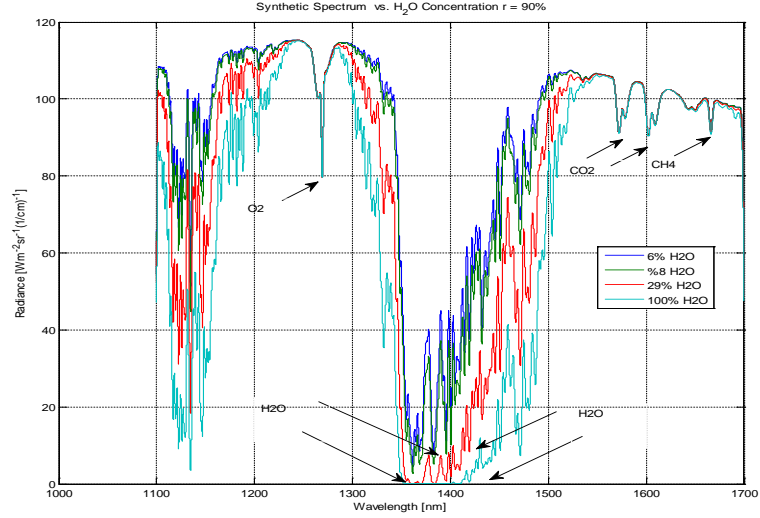

**Figure 7.** GENSPECT - synthetic spectrums with selected H$_2$O level concentration and albedo = 0.9.

## 3. Radiance enhancement approach

In order to analyse the spectral radiance data, we have introduced the RE and the Combined radiance enhancement formulae (CRE) (Siddiqui et al., 2015; Siddiqui, 2017)

$$RE_i = \frac{1}{N} \sum_{j=1}^{N} \left\{ \frac{OBS_i[j] - SYN_i[j]}{SYN_i[j]} \right\},$$

where $RE_i$ is the RE for all wavelength bands from $i$ to $j$, $OBS_i[j]$ is observed (real) data, $SYN_i[j]$ is synthetic (GENSPECT-simulated) data

$$CRE(1) = \frac{RE(2)}{RE(1) + RE(4)},$$

$$CRE(2) = \frac{RE(1)}{RE(2) + RE(3)},$$

$$CRE(3) = RE(1) + RE(2) + RE(4),$$

such that $CRE(1)$ is the combined RE (for $i = 1$ to $4$), $RE(1)$ is the RE in the H$_2$O and O$_2$ bands of interest, $RE(2)$ is the RE in the H$_2$O and CO$_2$ band of interest, $RE(3)$ is RE in the CO$_2$ band of interest, $RE(4)$ is RE in the CH$_4$ band of interest. Such a quantification of RE enables us to efficiently specify different areas for cloud scenes. Particularly, by comparing RE and CRE with corresponding Argus geo-locations with existing 2D cloud images from MODIS datasets, we can obtain the threshold values that can be used further to predict cloud scenes. As an example, this threshold value is shown in the Figure 21 by red vertical line.

The RE method is absolutely different as compared to the traditional methods of cloud detection. For example, paper (Wind et al., 2010) reported a classical way of cloud detection that provides large capabilities for precise and reliable probabilistic approach in determination of the multi-layered cloud masking. Although our method of cloud detection is unconventional, it utilises inexpensive, light and small-size remote sensing device. Consequently, the RE method is cost-efficient. Furthermore, as it works in the spectral range 1100 nm to 1700 nm that manifests profound spectral radiance for greenhouse gases including $CO_2$, $CH_4$, CO and $H_2O$ (Kuze et al., 2009; Tricoli, 2015). Consequently, it may be utilised simultaneously for continuous monitoring of the greenhouse gases that appears due to anthropogenic activities (Jagpal, 2011; Jagpal et al., 2019). Another significant advantage is that monitoring of the Earth-orbiting Argus instrument is periodic and covers the global areas repetitively along its track.

In fact, Argus instrument collects spectral radiance data along its path. Therefore, from this point of view its data is rather one dimensional. However, the cluster of the light, small-size and inexpensive Argus-like spectrometers operating in parallel in a network configuration, proposed by one of our co-authors in the work (Jagpal, 2011), can be used to provide datasets that may be suitable for 2D image processing. Such a technique employing RE method may have more efficient perspective for cloud scene detection due to larger coverage and capability for the IR 2D image processing. Particularly, this type of 2D imaging may be useful for analysis and visual interpretation of the cloud scenes.

The disadvantages of this method are related to limitation in spatial resolution (~1.4 km). As a result, the cloud scene boundary cannot be determined precisely. The RE method also cannot be used to classify types of the clouds (stratus, cumulus, stratocumulus, altostratus, altocumulus, cirrus, cirrocumulus, cumulonimbus, etc.) and unlike the method described in Wind et al. (2010), it cannot be used for multi-layer cloud masks. Moreover, this method does not distinguish clouds with aerosol pollutants (dust storm, volcano activities, forest fire smokes), heavy snow and ice since this substances increase surface albedo that causes very high solar light reflection. Despite that, the RE method may have a potential to detect seasonal forest fires that is especially relevant to North

Pacific and North Atlantic regions of Canada where massive areas are covered by forest (McCoy and Burn, 2005; Skinner et al., 2006; Wang et al., 2017). For example, if RE method indicates a 'cloud' scene that is not confirmed by a local weather forecast (as well as by real or near real-time MODIS or Google cloud database), then it is most likely to be smoke-originated aerosols (Diamond et al., 2018) appearing as a result of intensive combustion in the forest fire (Wiggins et al., 2018). As we can see, the RE method may find its application not only in detection of cloud scenes, but also in detection of smoke-originated aerosols due to seasonal forest fires (in Canada – from April to September (Wang et al., 2017)). Once such an event is detected timely, this information may help prevent spreading of fire over large areas in forests. Since this observation is repetitive due to cyclic rotation of the satellite in space, this technique enables us to monitor continuously the spreading or extinguishing dynamics of the forest fire.

Unlike the method comprehensively described in Wind et al. (2010), our RE method does not quantify probabilistic detection at the current stage of the project. Even though the probabilistic quantification would also be desirable in this approach, RE method requires a specific consideration that differs from that of shown in Wind et al. (2010). The probability metrics in detection of clouds may be developed in our future studies. Current ly, we use four-color gradation scheme for interpretation of existence of clouds (see Figure 8–16 in Siddiqui (2017) such that red, blue, green and grey colours signifies high, moderate, low and lowest chances for cloud existence. The albedo from the ground is one of the main key factors in the RE method that always is accounted for detection of the cloud scenes (see Figures 11–14, 18–26 in Siddiqui (2017) indicating the surface reflection values for each specific cases). Efficient detection of clouds or its surface features means detecting and recording of enhancement of radiant energy by clouds and their surface configuration at the border air–cloud. The detection of cloud or non-cloud scene is implemented by finding the maximum or minimum RE within selected NIR wavelength bands of $O_2$, $H_2O$, $CO_2$ and $CH_4$ (Siddiqui et al., 2015). The RE approach is mainly based on the mean value of the ratio of the difference of the observed data with simulated data for the selected week per pass with single scan or multiple scan (Siddiqui, 2017). In this model the cloud detection can found by selecting the sun elevation angle, satellite nadir angle, variable path length, atmospheric water vapour, variable reflectance, and cloud structure over land or sea (Siddiqui, 2017). Table 3 shows the input parameters used for the efficient detection of cloud scene with geolocation of the Argus flight.

**Table 3.** Input parameters for RE model.

| Types of parameter | Significance values and ranges |
|---|---|
| Mixing Ratios of gases | $O_2$.mxr, $CO_2$.mxr, $CH_4$.mxr, refmod 95_ $H_2O$.mxr (1976 U.S. Standard Atmospheric Model) |
| Gases in % | $O_2$ (100) , $CO_2$ (100), $CH_4$ (100), $H_2O$ (0 to 35) |
| Height from surface to top of clouds | 2km to 50 km |
| Surface Type | Lambertian |
| Reflectivity | 0.3 (over generic vegetation and bare soil) 01 to 0.9 (over snow, clouds, and ice) |
| Scattering Type | Rayleigh |

For the analysis of efficient detection of cloud scenes by using RE approach we have selected few retrieval datasets from Argus flight as shown in Table 2. The RE approach for the selected GENSPECT-synthetic model with albedo 0.3 at different water level concentrations in contrast with Argus selected spectra of different week per pass per observation numbers are shown in Figures 8, 9 and 10.

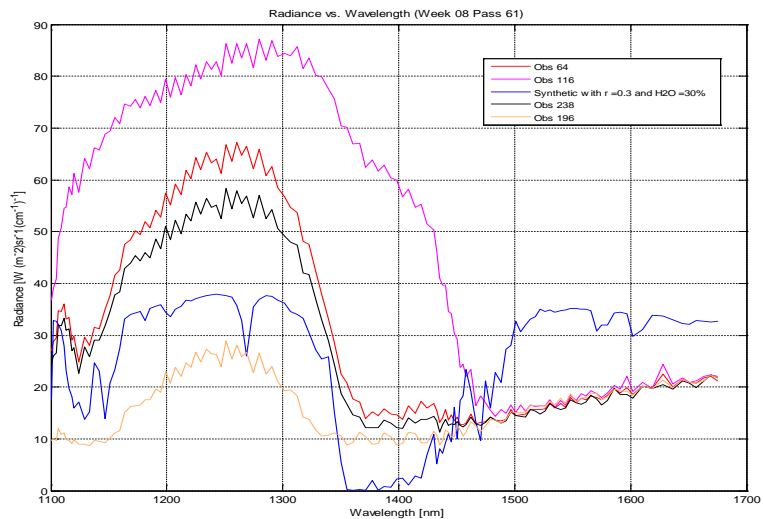

**Figure 8.** RE for Argus spectra in clear and cloudy sky of week 08 pass 61 with different observations number 64/116/196/238 vs. synthetic model spectra with (r = 0.3, $H_2O$ = 30%).

Figures 8, 9 and 10 show the different selected observation numbers of weeks 08, 14 and 75 and passes 61, 52 and 43. Each observation number has been compared with synthetic model with albedo 0.3 and $H_2O$ concentration as 10% and 30%. The observation numbers 64, 116 and 238 of week 08 pass 61 and observation numbers 22 and 124 of week 14 pass 52 ranked as cloud signature. Similarly observation number 196 of week 08 pass 61, observations number 100 and 120 of week 14 pass 52 and observations numbers 19, 30, 43 and 65 of week 75 pass 43 ranked as non-cloud scene by using RE values as shown in Table 4.

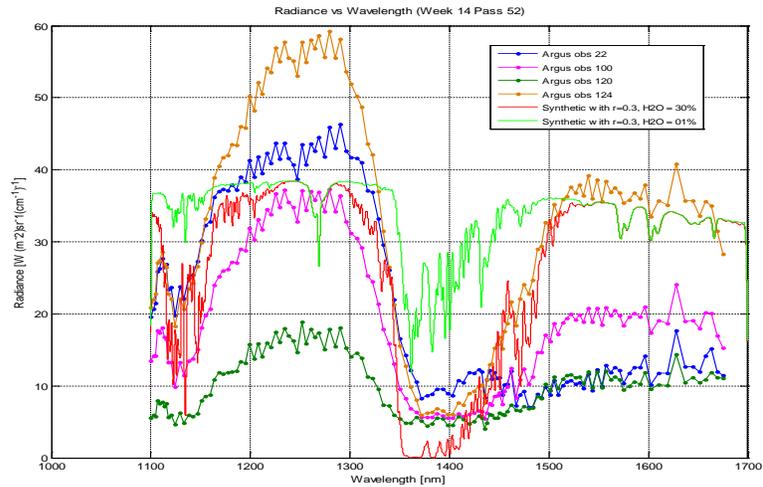

**Figure 9.** RE for Argus spectra in clear and cloudy sky of week 14 pass 52 with different observations number 22/100/120/124 vs. synthetic model spectra with (r = 0.3, $H_2O$ = 1% & 30%).

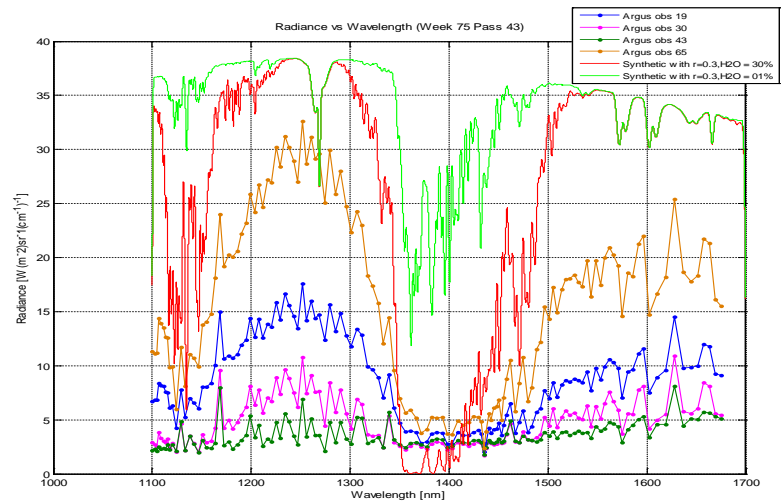

**Figure 10.** RE for Argus spectra in clear sky of week 75 pass 43 with different observations number 19/30/43/65 vs. synthetic model spectra with (r = 0.3, H2O = 1% and 30%).

## 4. Shortwave upwelling radiative flux approach

The integrated absorption technique is applied to develop a synthetic model to determine the magnitude of ShortWave upwelling Radiative Flux (SWupRF) within NIR wavelength bands of $O_2$, $H_2O$, $CO_2$ and $CH_4$ (Siddiqui et al., 2017, Siddiqui, 2017). This new synthetic model is used to estimate the magnitude and expected magnitude variation over spectral range of 900 nm to 1700 nm by varying surface temperature to assess effect on outgoing (upwelling) forcing term (Siddiqui et al., 2016a). In this approach, we employ satellite real observation of space orbiting Argus 1000 for $O_2$, $H_2O$, $CO_2$ and $CH_4$ with all the packets of the specified weeks, calibration, and background files to calculate the SWupRF. The SWupRF model loads a set of observed spectra for different week per pass per observation number and integrates each spectrum over the different spectral range of bands of interest to compute the (SWupRF)$_{obs}$ (W/m$^2$) as shown in Figures 11, 12 and 13.

Figures 11, 12 and 13 show the SWupRF of the selected observation numbers of weeks 08, 14 and 75 and passes 61, 52 and 43. The higher the flux (W/m2) of different observations of selected week per pass of Argus flight higher the chances of cloud scene. The lower flux profile at different observations demonstrates the clear sky or patches of clouds. The results of SWupRF are shown in Table 6.

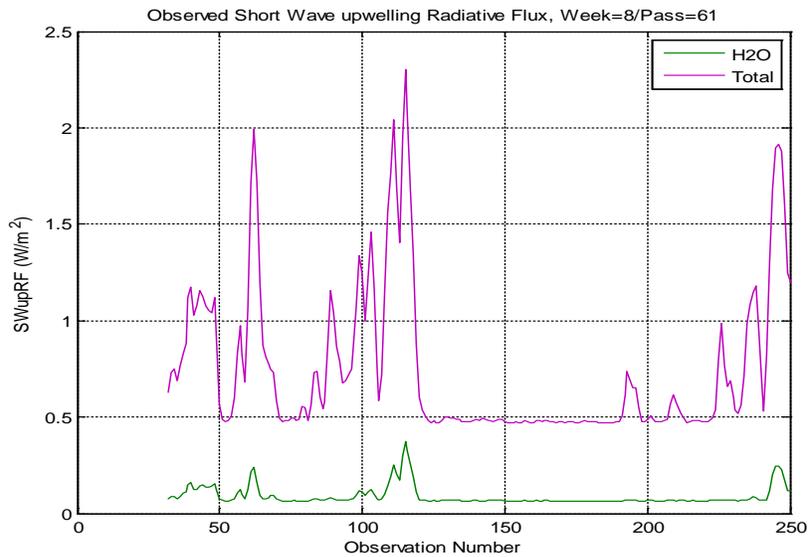

**Figure 11.** SWupRF of week 08 pass 61 of Argus observed data.

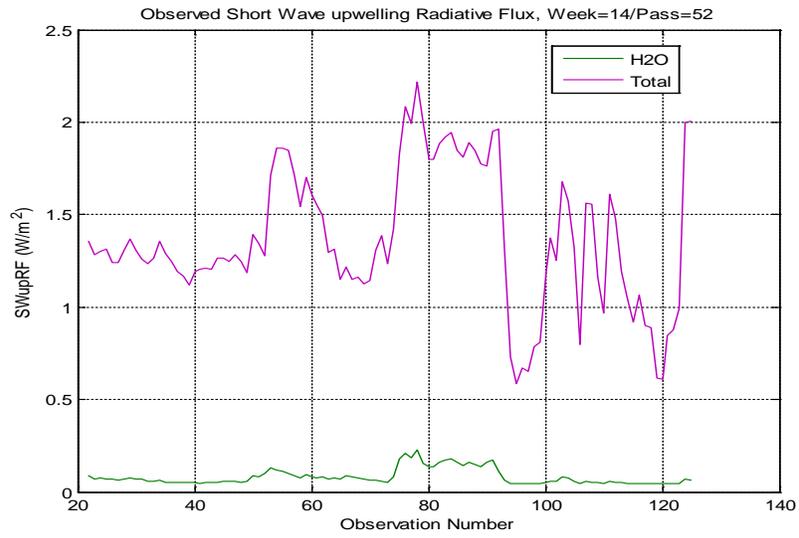

**Figure 12.** SWupRF of week 14 pass 52 of Argus observed data.

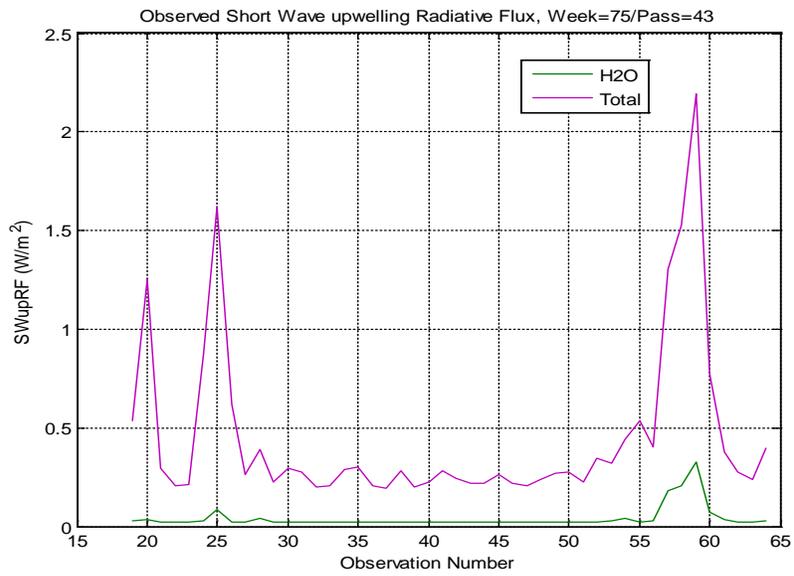

**Figure 13.** SWupRF of week 75 pass 43 of Argus observed data.

# 5. Validations of cloud and non-cloud scenes

The performance of efficient detection of cloud scenes is usually based on by associations to the other satellites imagery (Jedlovec, 2009). In this study the validations have been carried out on the basis of MODIS-Aqua/Terra satellite imageries (see http://modis-atmos.gsfc.nasa.gov/IMAGES/index.html). Figures 14, 15 and 16 (Siddiqui, 2017) present the agreement of RE and SWupRF based results of Argus data set and the MODIS cloud images. The validation is splits into three types of scenarios of cloud scenes showing the variability of different types of cloud surface intensity.

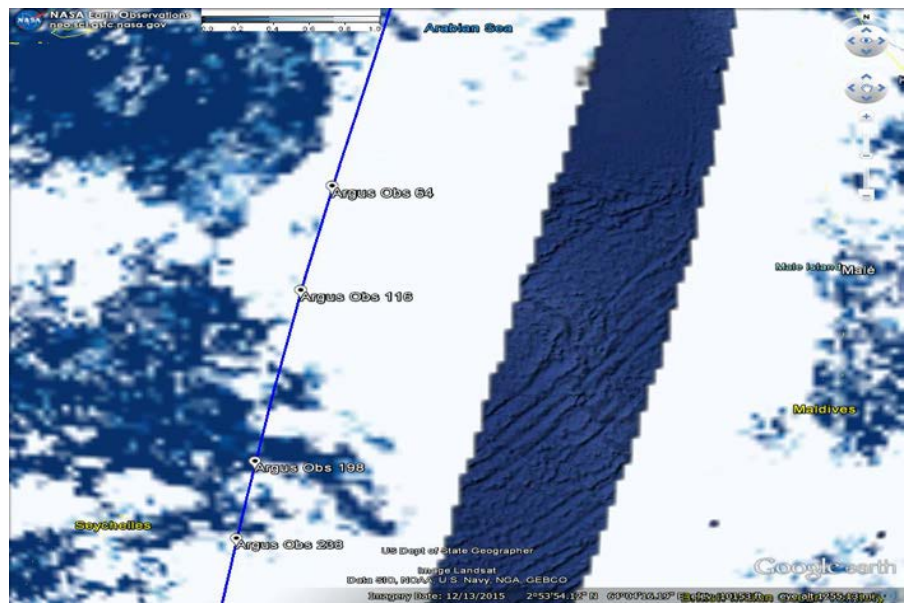

**Figure 14.** Argus 1000 infrared space flight path with MODIS cloud masks of week 08 pass 61 for October 30, 2009 over Arabian Sea.

It is very important to compare the inside features of numerous clouds detection methodology because they are often used in different settings. Each cloud detection method may use different satellite sensors, different wavelength selections bands, and different geographical regions, different date and time etc. (Jedlovec, 2009). In our analysis all the three selected data set of Argus space flight showing a very good agreement in performance to detect efficiently the cloud scenes over different regions around the globe in comparisons with MODIS – Aqua/Terra satellite cloud masks. A quantitative analysis of all the selected data set of Argus 1000, as shown in Table 4,indicates that all the validations of cloud scenes ensured reasonably well in capturing the clouds and non-cloud scenes by the results of both RE and SWupRF techniques.

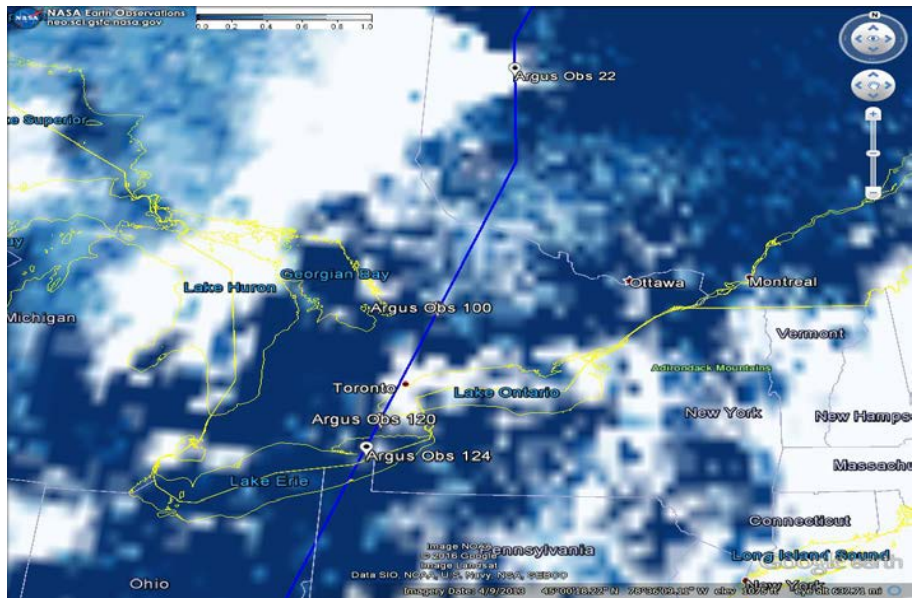

**Figure 15.** Argus 1000 infrared space flight path with MODIS cloud masks of week 14 pass 52 for March 04, 2010 over Ontario, Canada.

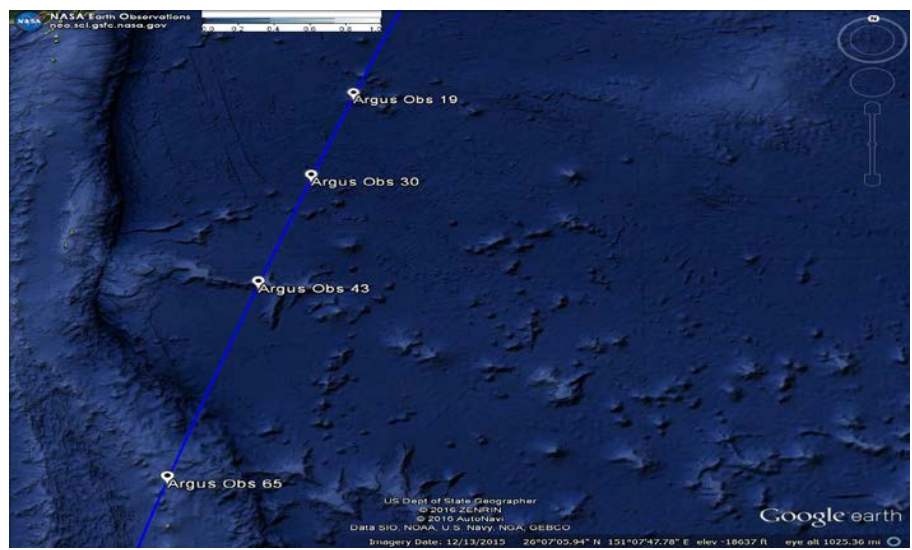

**Figure 16.** Argus 1000 infrared space flight path with MODIS no-cloud masks of week 75 pass 43 for August 14, 2013 over North Pacific Ocean.

**Table 4.** Argus selected week per pass per observations with geo-location

| W/P/O | Albedo | Altitude (Km) | Radiance (Max.) W/m²sr⁻¹(cm⁻¹)⁻¹ | Radiance (Min.) W/m²sr⁻¹(cm⁻¹)⁻¹ | RE | SWupRF (W/m²) | Cloud/Type |
|-------|--------|---------------|------------------|------------------|------|--------|------------|
| 08/61/64 | 0.5 | 10 | 83 | 17 | 9.86 | 2.1 | Yes/Thick (full) |
| 08/61/196 | 0.2 | 10 | 30 | 09 | 4.90 | 0.8 | Yes/Thin (partial) |
| 08/61/238 | 0.4 | 10 | 40 | 09 | 7.91 | 1.2 | Yes/Thick (full) |
| 14/52/22 | 0.5 | 10 | 60 | 12 | 4.70 | 1.3 | Yes/Thin (full) |
| 14/52/100 | 0.3 | 10 | 36 | 06 | 6.13 | 1.2 | Yes/Thin (partial) |
| 14/52/125 | 0.5 | 10 | 60 | 06 | 2.13 | 2.0 | Yes/Thin (partial) |
| 75/43/30 | 0.02 | 02 | 06 | 0.5 | -1.53 | 0.25 | No/nil (ocean) |
| 75/43/43 | 0.1 | 02 | 13 | 03 | -1.52 | 0.25 | No/nil (ocean) |
| 75/43/65 | 0.2 | 05 | 30 | 04 | 1.20 | 1.1 | No/nil (surface ice) |

# 6. Results and discussions

The results are assembled in Figures 17, 18 and 19 to show snapshots for potential candidates of cloud non-cloud scenes from different Argus flight weeks per passes per observation by using RE & SWupRF models.

Figures 20 and 21 (Siddiqui, 2017) show the histograms of the obtained values of subsequent probability of cloud and non-cloud scenes. Both the results of RE and SWupRF for week 08 pass 61 with observation number 37 and week 75 pass 43 with observation number 115 and 116 are agreed for the efficient detection of cloud scenes. Higher the RE for the full wavelength bands as well as for the wavelength bands of $H_2O$ in contrast with the high intensity (W/m²) signifies the more probability of cloud scenes.

Figure 22 illustrates a correlation between RE and SWupRF model by using scattered plot with linear fitting to check the validity of the both models for cloud and non-cloud scenes. The scatter plot shows that most points are distributed near the 0.2–1.0 (W/m²): 1–10 (RE), which is most probability of non-clouds scenes (or due to the reflection of aerosols with cluster of dust particles). Figure 22 also illustrates that the higher the flux intensity in contrast with the RE that signifies the higher probability of cloud scenes. The overall results of RE and SWupRF (W/m$^2$) show an excellent commitment for the efficient detection of cloud scenes. There is some degree of data diversions because of number of errors, i.e.:

1) difference between satellite path and Argus boresight,
2) mixing of water vapor and carbon dioxide within atmospheric layers that affects calculation of the RE,
3) selection of average number of satellite sun and nadir angle while comparing with each set of Argus data set.

Higher values indicates that a cloud reflects a large amount of solar radiation by different types of clouds, this can be justified by the variation of cloud albedo from 10% to more than 90%, with different concentration of liquid water, thickness of the atmospheric layers (by changing altitudes), and the satellite sun's and nadir angle. The smaller the droplets signify the greater liquid water content and the greater cloud albedo under assumption that all other factors are the same for using detection of the clouds (Kramer, 2002).

The scattered plot as illustrated in Figure 22 gives the probability of different regions of cloud statistics. The higher the radiance with albedo from 0.7 to 0.9 with high altitudes, more the chances of low, thick clouds (such as stratocumulus) primarily reflecting most of the incoming solar radiation whereas with low albedo from 0.1 to 0.6 with high and low altitudes, more the chances of high, thin clouds (such as Cirrus) that tend to transmit it to the surface but then trap outgoing infrared radiation because of low albedo (Siddiqui, 2017).

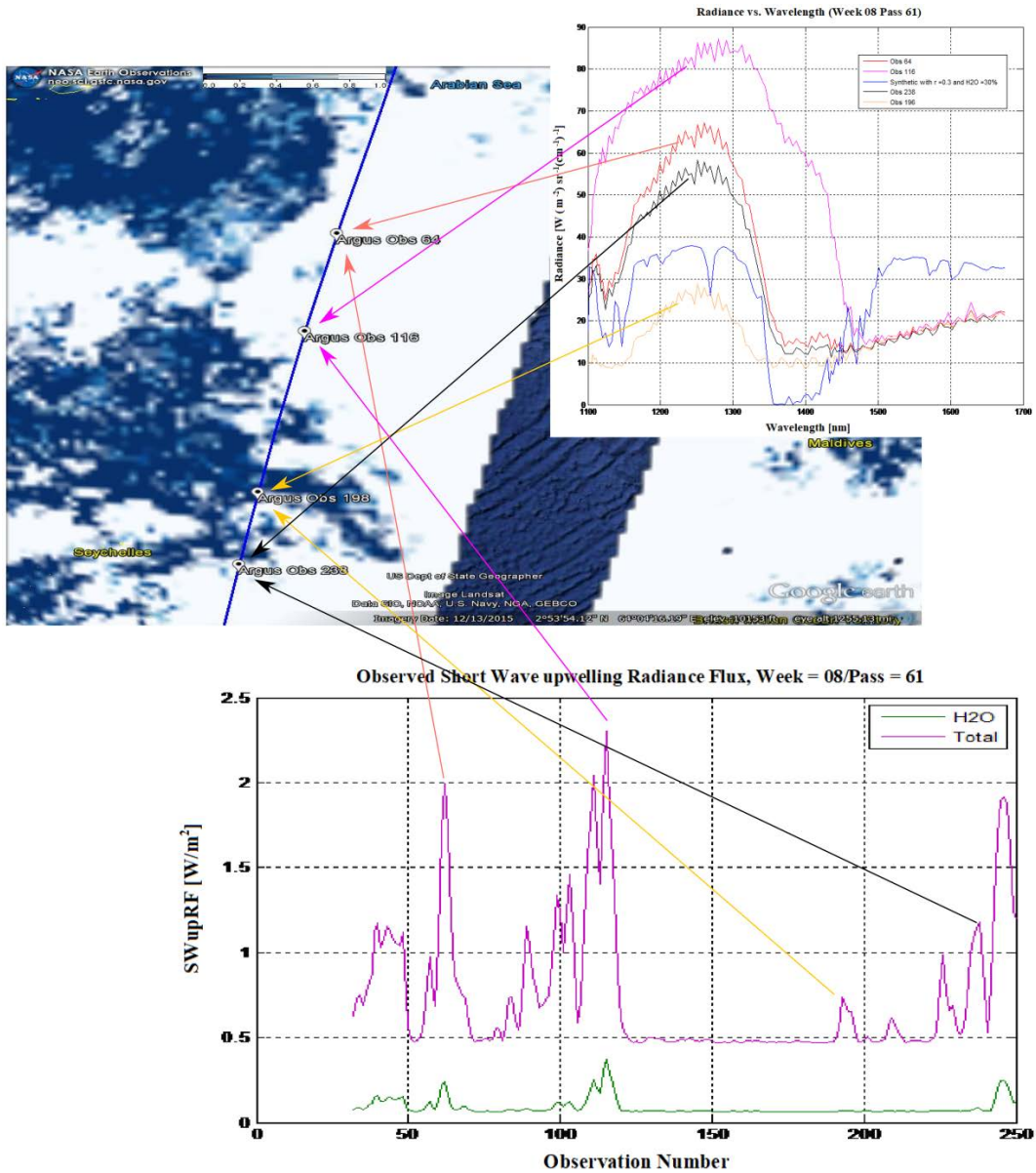

**Figure 17.** (a) RE for Argus week 08 pass 61 with observations number 64/116/198/238 vs. GENSPECT-synthetic model. (b) Argus flight vs. Terra/Aqua (MODIS cloud sat) with full and partial cloud scenes over Arabian Sea & Seychelles  (c) SWupRF (0.47- 2.30 W/m$^2$) shows the high and low radiative flux intensity within the same range of Argus observation number.

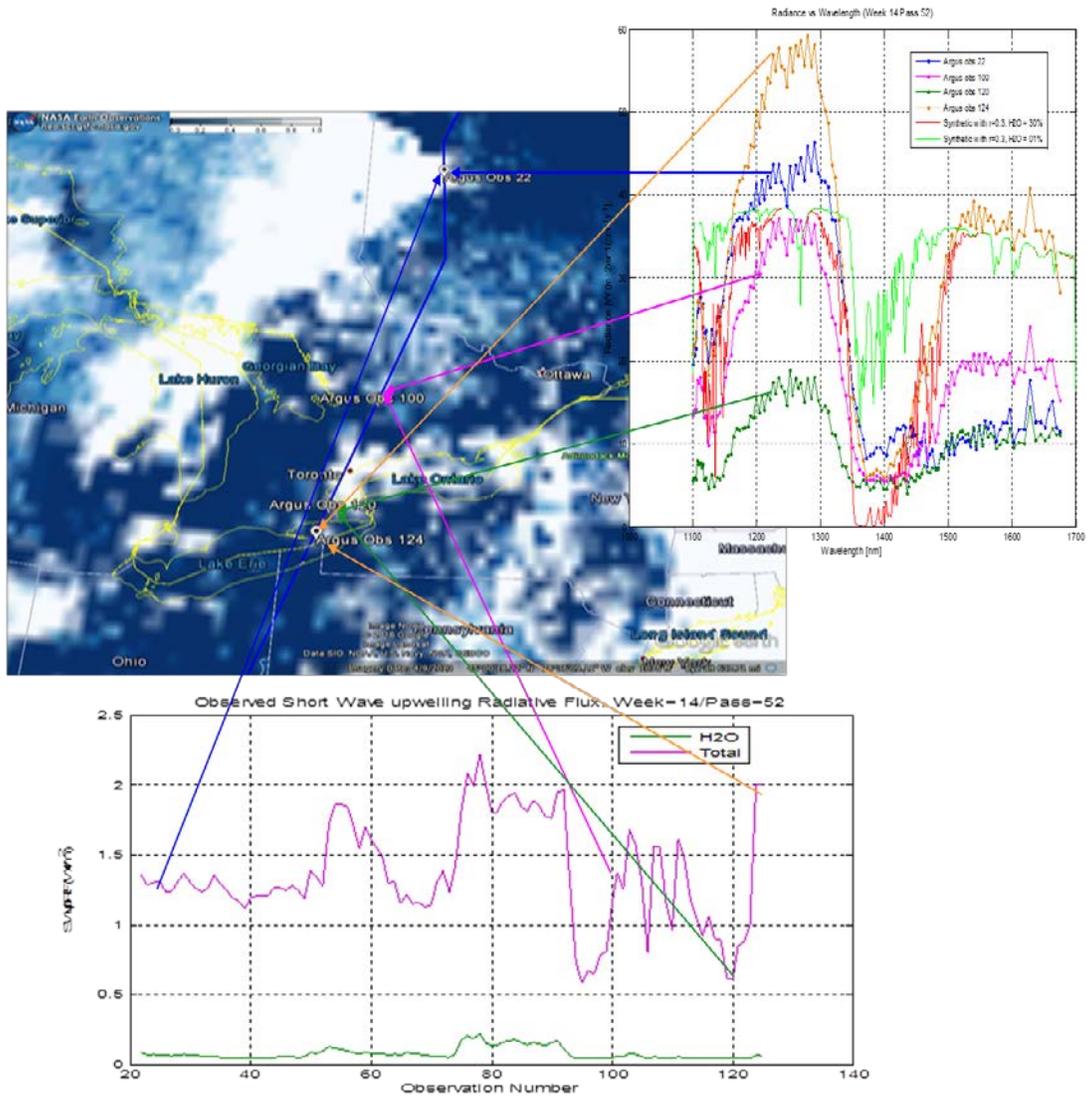

**Figure 18.** (a) RE for Argus week 14 pass 52 with observations number 22/100/120/124 vs. GENSPECT-Synthetic model. (b) Argus flight vs. Terra/Aqua (MODIS cloud sat) with full and partial cloud scenes over Ontario, Canada (c) SWupRF (0.60- 2.20 W/m$^2$) shows the high and low radiative flux intensity within the same range of Argus observation number.

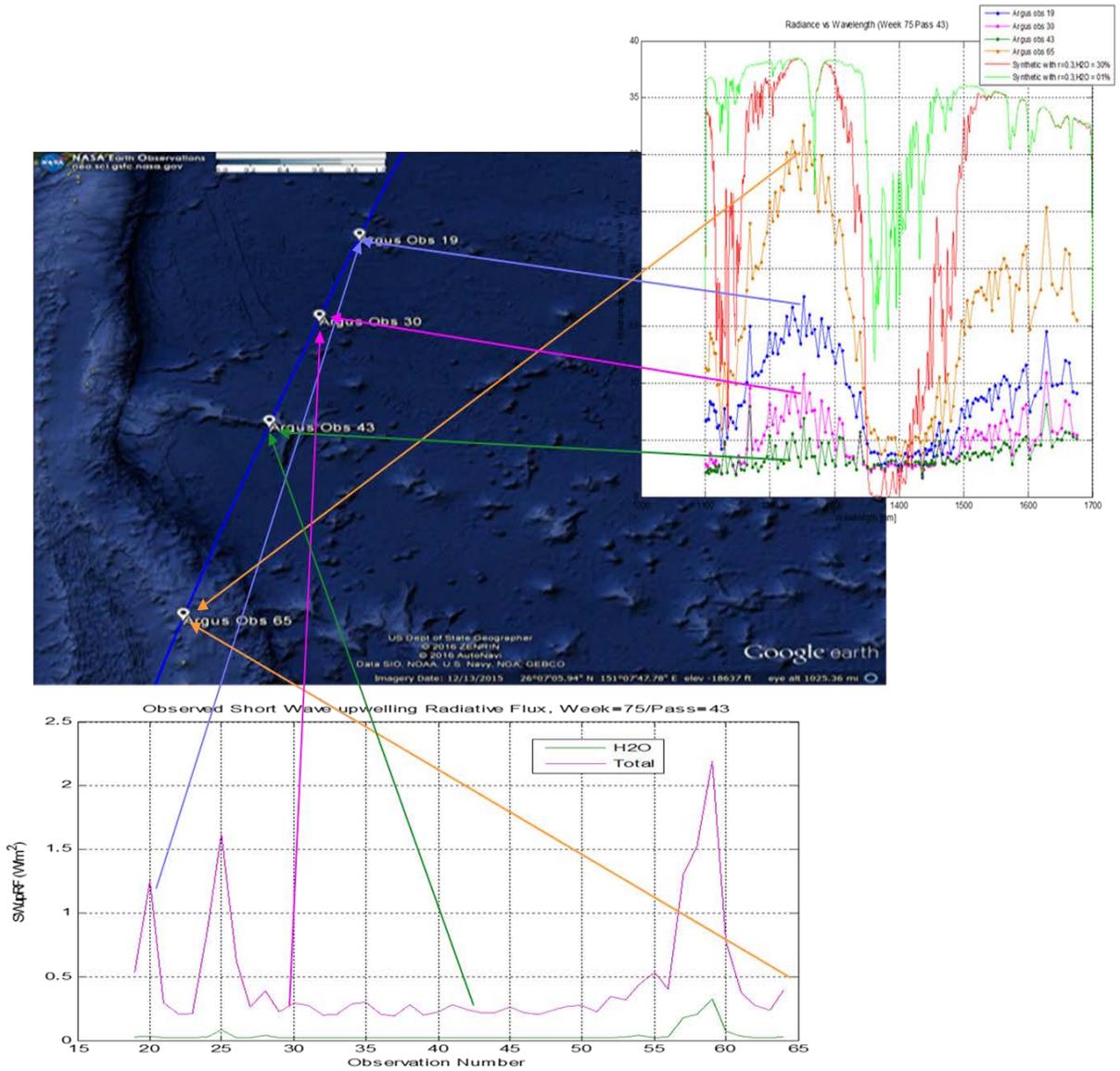

**Figure 19.** (a) RE for Argus week 75 pass 43 with observations number 19/30/43/65 vs. GENSPECT-synthetic model. (b) Argus flight vs. Terra/Aqua (MODIS cloud sat) with full and partial cloud scenes over North Pacific Ocean (c) SWupRF (0.19- 2.20 W/m$^2$) shows the high and low radiative flux intensity within the same range of Argus observation number.

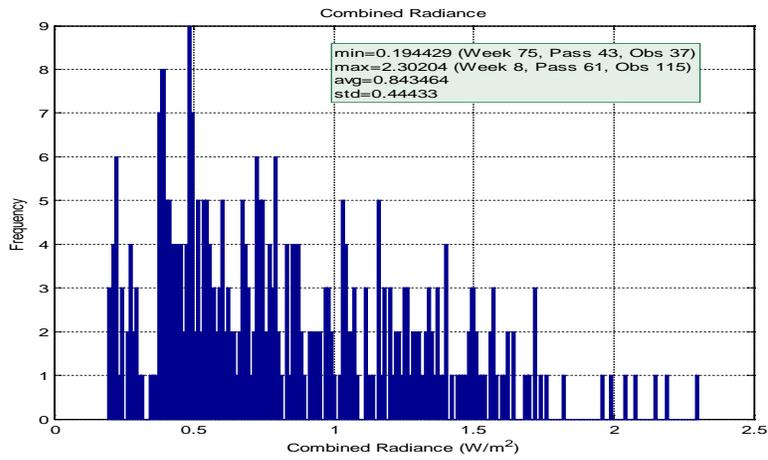

**Figure 20.** Histogram of the Argus weeks/passes/observations with maximum flux intensity = 2.30 W/m$^2$, minimum flux intensity = 0.2 W/m$^2$, average of full spectral data set = 0.84 W/m$^2$.

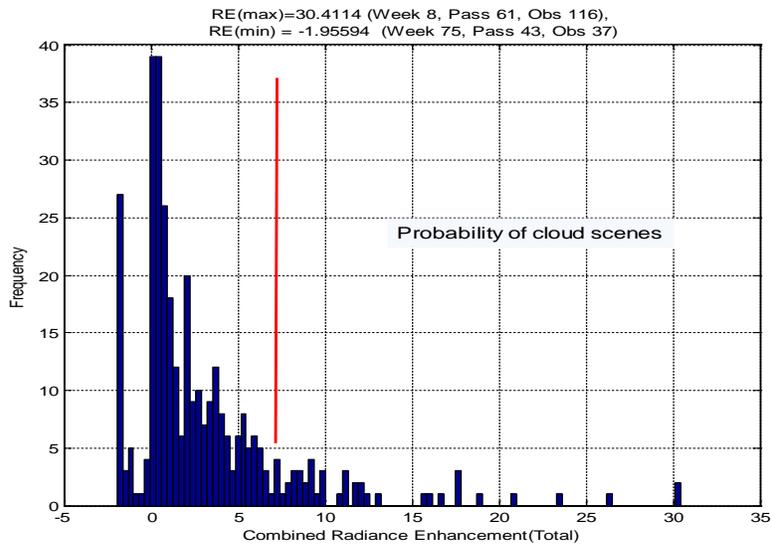

**Figure 21.** Histogram of the subsequent probability of cloud and non-cloud scenes.

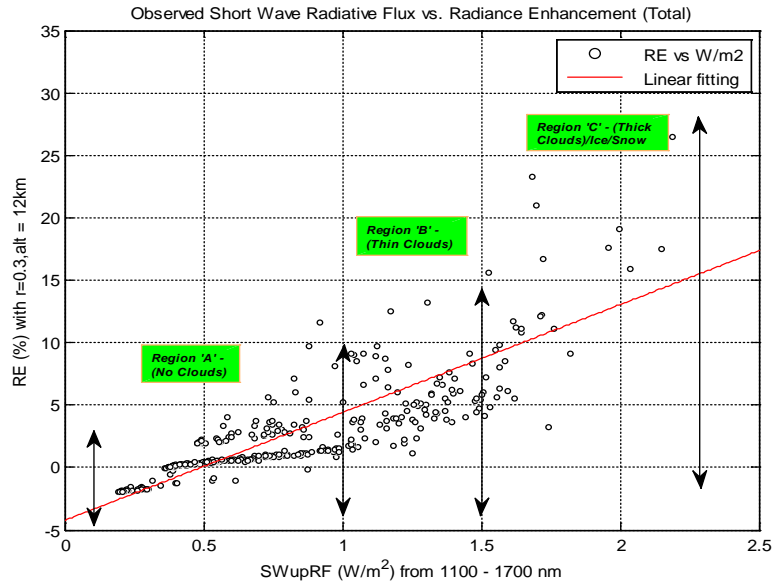

**Figure 22.** Scattered plot between RE and SWupRF of Argus data set.

## 6. Conclusions

In this work, we have applied two new approaches, RE and Shortwave upwelling Radiative Flux (SWupRF) approaches within NIR spectral range from 1100 nm to 1700 nm of space orbiting Argus micro-spectrometer over different spatial locations since 2009 for an improved cloud detection scenes. The two methods have been validated by real observations using collected MODIS data imagery. The RE-based method allows the efficient detection of the clouds through their higher spatial values in contrast with the GENSPECT line-by-line radiative transfer model within the high spectral resolution. This method mostly enables us to calculate the enhancement of reflectivity by different atmospheric concentrations of $CO_2$ and $H_2O$, range of albedos, satellite sun and nadir angle.

The second method tested is based on the SWupRF analysis within the same cluster of calculated radiances by the instrument. The cloud detection technique is applied on the high values of radiative flux intensity in terms of $W/m^2$ range of integrated spectral profile for the selected range of real observations as well in contrast with the RE results. Both results are in a good agreement with efficient detection cloud and non-cloud scenes by Argus FOV. Cloud detection at night is more challenging with described infrared measurements. The comprehensive investigation has also been required to add full range of Argus geo-located dataset. The presented methodology can reduce the quantification process of detection of cloud scenes and its relationships with the different atmospheric mixing ratios concentration, which is actively participated for the formation of clouds and will be helpful for the description of the climate change mechanism. The RE approach may also be promising in detection of smoke-originated aerosols appearing as a results of the seasonal forest fires.

## Acknowledgments

This study is supported by Department of Physics and Astronomy at York University, Epic College of Technology, Epic Climate Green (ECG) Inc., Canadian Advanced Nano space Experiment 2 Operations Team and Thoth Technologies, Inc. The authors would like to thanks to all of the organisations for their support, guidance and suggestions in operating CanX-2 spacecraft (Rankin, 2005; Sarda et al., 2006). Useful comments and constructive discussions from Harshal Gunde (Epic Climate Green (ECG) Inc.) are greatly appreciated. Special thanks are addressed to Dr. Robert E. Zee (University of Toronto Institute for Aerospace Studies Space) for providing Argus 1000 space datasets from CanX-2 spacecraft.

## References

Abramowitz, M. and Stegun, I.A. (1972) 'Error function and Fresnel integrals', *Handbook of Mathematical Functions with Formulas, Graphs, and Mathematical Tables*, National Bureau of Standards, 9th ed., New York.

Abrarov, S.M., Quine, B.M. and Jagpal, R.K. (2010a) 'Rapidly convergent series for high-accuracy calculation of the Voigt function', *Journal of Quantitative Spectroscopy and Radiative Transfer*, Vol. 111, No. 3, pp.372–375.

Abrarov, S.M., Quine, B.M. and Jagpal, R.K. (2010b) 'High-accuracy approximation of the complex probability function by Fourier expansion of exponential multiplier', Computer *Physics Communication*, Vol. 181, No. 5, pp.876–882.

Ackerman, S.A., Strabala, K.I., Menzel, W.P., Frey, R.A., Moeller, C.C. and Gumley, L.E. (1998) 'Discriminating clear sky from clouds with MODIS', *Journal of Geophysical Research*, Vol. 103, No. D24, pp.32–141.

Chesser, H., Lee, R., Benari, G., Jagpal, R., Lam, K. and Quine, B. (2012) 'Geolocation of Argus flight data', *IEEE Transactions on Geoscience and Remote Sensing*, Vol. 50, No. 2, pp.357–361.

Christopher, S.A. and Gupta, P. (2010) 'Satellite remote sensing of particulate matter air quality: the cloud-cover problem', *Journal of the Air and Waste Management Association*, Vol. 60, No. 5, pp.596–602.

Cutillo, L., Amato, U., Antoniadis, A., Cuomo, V. and Serio, C. (2004) 'Cloud detection from multispectral satellite images', *Proceedings from IEEE Gold Conference*, University Parthenope, Naples, Italy.

de Leeuw, G., Kokhanovsky, A.A. and Cermak, J. (2012) 'Remote sensing of aerosols and clouds: Techniques and applications (editorial to special issue in Atmospheric Research)'. *Atmospheric Research*, Vol. 113, pp.40–42.

Diamond, M.S., Dobracki, A., Freitag, S., Griswold, J.D.S., Heikkila, A., Howell, S.G., Kacarab, M.E., Podolske, J.R., Saide, P.E. and Wood, R. (2018) 'Time-dependent entrainment of smoke presents an observational challenge for assessing aerosol – cloud interactions over the southeast Atlantic Ocean', *Atmospheric Chemistry Physics*, Vol. 18, pp.14623–14636.

Duggan, P., Sinclair, P.M., LeFlohic, M.P., Forsman, J.W., Berman, R., May, A.D. and Drummond, J.R. (1993) 'Testing the validity of the optical diffusion coefficient: line-shape measurements of CO perturbed by $N_2$', *Physical Review A*, Vol. 48, No. 5, pp.2077–2083.

Ebert, E.E. (1987) 'A pattern recognition technique for distinguishing surface and cloud types in the polar regions', *Journal of Climate and Applied Meteorology*, Vol. 26, No. 10, pp.1412–1427.

Fournier, N., Stammes, P., Graaf, M.D., Piters, A., Grzegorski, M. and Kokhanovsky, A. (2006) 'Improving cloud information over deserts from SCIAMACHY oxygen A-band measurements', *Atmospheric Chemistry Physics*, Vol. 6, No. 1, pp.163–172.

Ghosh, R.R., Ali, M.S., Hena, A. and Rahman, H. (2012) 'A simple cloud detection algorithm using NOAA-AVHRR satellite data', *International Journal of Science and Engineering Research*, Vol. 3, No. 6, pp.1–5.

Guo, F., Shen, X., Zou, L., Ren, Y., Qin, Y., Wang, X. and Wu, J. (2015) 'Cloud detection method based on spectral area ratios in MODIS data', Canadian Journal of Remote Sensing, Vol. 41, No. 6, pp.561–576.

Hatzianastassiou, N., Matsoukas, C., Fotiadi, A., Pavlakis, K.G., Drakakis, E., Hatzidimitriou, D. and Vardavas, I. (2005) 'Global distribution of Earth's surface shortwave radiation budget', *Atmospheric Chemistry Physics*, Vol. 5, No. 10, pp.2847–2867.

Illingworth, A.J., Barker, H.W., Beljaars, A., Ceccaldi, M., Chepfer, H., Clerbaux, N., Cole, J., Delanoë, J., Domenech, C., Donovan, D.P. and Fukuda, S. (2015) 'The EarthCARE satellite: the next step forward in global measurements of clouds, aerosols, precipitation, and radiation', *Bulletin of American Meteorological Society*, Vol. 96, No. 8, pp.1311–1332.

Inamdar, A.K. and Guillevic, P.C. (2015) 'Net surface shortwave radiation from GOES imagery – product evaluation using ground-based measurements from SURFRAD', *Remote Sensing*, Vol. 7, No. 8, pp.10788–10814.

Jagpal, R.K. (2011) Calibration and Validation of Argus 1000 Spectrometer – a Canadian Pollution Monitor, PhD Thesis, York University, Canada.

Jagpal, R.K., Quine, B.M., Chesser, H., Abrarov, S.M. and Lee, R. (2010) 'Calibration and in-orbit performance of the Argus 1000 spectrometer-the Canadian pollution monitor', *Journal of Applied Remote Sensing*, Vol. 4, No. 1, p.049501.

Jagpal, R.K., Siddiqui, R., Abrarov, S.M. and Quine, B.M. (2019) 'Carbon dioxide retrieval of Argus 1000 space data by using GENSPECT line-by-line radiative transfer model', *Environment Natural Resources Research*, Vol. 9, No. 3, pp.77–85.

Jedlovec, G. (2009) 'Automated detection of clouds in satellite imagery', *Advances in Geoscience and Remote Sensing*, Jedlovec, G. (Ed.), InTech.

Kramer, H.J. (2002) *Observation of the Earth and its Environment: Survey of Missions and Sensors*, Springer Science & Business Media.

Kuze, A., Suto, H., Nakajima, M. and Hamazaki, T. (2009) 'Thermal and near infrared sensor for carbon observation Fourier-transform spectrometer on the greenhouse gases observing satellite for greenhouse gases monitoring', *Applied Optics*, Vol. 48, No. 35, pp.6716–6733.

Li, W. and Li, D. (2008) 'Cloud detection in MODIS data based on spectrum analysis and snake model', in *Geoinformatics 2008 and Joint Conference on GIS and Built Environment: Classification of Remote Sensing Images*, p.714710, International Society for Optics and Photonics.

Li, Z. and Leighton, H.G. (1991) 'Scene identification and its effect on cloud radiative forcing', *Journal of Geophysical Research*, Vol. 96, pp.9175–9188.

Mao, J. and Kawa, S.R. (2004) 'Measurement of atmospheric carbon dioxide column from space using reflected sunlight', in *Remote Sensing*, pp.336–346, International Society for Optics and Photonics, Barcelona, Spain.

McCoy, V.M. and Burn, C.R. (2005) 'Potential alteration by climate change of the forest-fire regime in the boreal forest of Central Yukon Territory', *Arctic*, Vol. 58, No. 3, pp.276–285.

McKernan, E., Quine, B.M. and Drummond, J.R. (2002) 'MOPITT sensitivity studies', Computation of instrument parameter dependencies, *2002 IEEE International Geoscience and Remote Sensing Symposium (IGARSS)*, Vol. 2, pp.1102–1104.

Mishchenko, M.I., Rossow, W.B., Macke, A. and Lacis, A.A. (1996) 'Sensitivity of cirrus cloud albedo, bidirectional reflectance and optical thickness retrieval accuracy to ice particle shape', *Journal of Geophysical Research: Atmospheres*, Vol. 101, D12, pp.16973–16985.

Mitchell, D.L. and Finnegan, W. (2009) 'Modification of cirrus clouds to reduce global warming', Environment Research Letters, Vol. 4, No. 4, p.045102.

MODIS [online] http://modis-atmos.gsfc.nasa.gov/IMAGES/index.html, Accessed on 2019-12-09.

Pagano, T.S. and Durham, R.M. (1993) 'Moderate resolution imaging spectroradiometer (MODIS)', in *Optical Engineering and Photonics in Aerospace Sensing*, pp.2–17, International Society for Optics and Photonics, Orlando, FL, United States.

Quine, B.M. and Drummond, J.R. (2002) 'GENSPECT: a line-by-line code with selectable interpolation error tolerance', Journal of Quantitative Spectroscopy and Radiative Transfer, Vol. 74, No. 2, pp.147–165.

Rankin, D., Kekez, D.D., Zee, R.E., Pranajaya, F.M., Foisy, D.G. and Beattie, A.M. (2005) 'The CanX-2 nanosatellite: expanding the science abilities of nanosatellites', *Acta Astronautica*, Vol. 57, Nos. 2–8, pp.167–174.

Rossow, W.B. and Garder, L.C. (1993) 'Cloud detection using satellite measurements of infrared and visible radiances for ISCCP', *Journal of Climate*, Vol. 6, No. 12, pp.2341–2369.

Sarda, K., Eagleson, S., Caillibot, E., Grant, C., Kekez, D., Pranajaya and F. and Zee, R.E. (2006) 'Canadian advanced nanospace experiment 2: scientific and technological innovation on a three-kilogram satellite', *Acta Astronautica*, Vol. 59, Nos. 1–5, pp.236–245.

Schiffer, R.A. and Rossow, W.B. (1983) 'The international satellite cloud climatology project (ISCCP) – the first project of the world climate research programme', *American Meteorological Society Bulletin*, Vol. 64, No. 7, pp.779–784.

Schreier, M. and Suselj, K. (2016) 'Analysis of collocated AIRS and MODIS data: a global investigation of correlations between clouds and atmosphere in 2004–2012', *International Journal of Remote Sensing*, Vol. 37, No. 11, pp.2524–2540.

Siddiqui, R. (2017) *Efficient detection of cloud scenes by a space-orbiting Argus 1000 micro-spectrometer*, PhD Thesis, York University, Toronto, Canada.

Siddiqui, R., Jagpal, R. and Quine, B.M. (2017) 'Short Wave upwelling Radiative Flux (SWupRF) within near infrared (NIR) wavelength bands of O2, H2O, CO2, and CH4 by Argus 1000 along with GENSPECT line by line radiative transfer model', *Canadian Journal of Remote Sensing*, Vol. 43, No. 4, pp.330–344.

Siddiqui, R., Jagpal, R., Salem, N.A. and Quine, B.M. (2015) 'Classification of cloud scenes by Argus spectral data', *International Journal of Space Science and Engineering*, Vol. 3, No. 4, pp.295–311.

Siddiqui, R., Jagpal, R., Salem, N.A. and Quine, B.M. (2016a) 'Short wave upwelling radiative flux (SWupRF) within NIR range for the selected greenhouse wavelength bands of $O_2$, $H_2O$, $CO_2$ and $CH_4$ by Argus 1000 along with GENSPECT line by line radiative transfer model', *arXiv Preprint*, arXiv: 1608.05386.

Siddiqui, R., Jagpal, R., Salem, N.A. and Quine, B.M. (2016b) 'The efficient detection of cloud scenes by radiance enhancement (RE) and with their impact on earth global energy budget due to short wave upwelling radiative flux (SWupRF) within NIR spectral range of space orbiting Argus1000 micro spectrometer', in *ExoClimes Conference Preceding*, NASA Astrobiology Institute [online] https://nai.nasa.gov/media/medialibrary/2016/08/Siddiqui_CloudDetectionREandSWupRF_Poster_Excolimes2016.pdf, Accessed on 2019-12-09.

Skinner, W.R., Shabbar, A., Flannigan, M.D. and Logan, K. (2006) 'Large forest fires in Canada and the relationship to global sea surface temperatures', *Journal of Geophysical Research*, Vol. 111, D14106.

Someya, Y., Imasu, R., Saitoh, N., Ota, Y. and Shiomi, K. (2016) 'A development of cloud top height retrieval using thermal infrared spectra observed with GOSAT and comparison with CALIPSO data', *Atmospheric Measurement Technique*, Vol. 9, No. 5, pp.1981–1992.

Song, X., Liu, Z. and Zhao, Y. (2004) 'Cloud detection and analysis of MODIS image', in *2004 IEEE International Proceedings Geoscience and Remote Sensing Symposium, IGARSS'04*, Vol. 4, pp.2764–2767.

Tang, H., Yu, K., Hagolle, O., Jiang, K., Geng, X. and Zhao, Y. (2013) 'A cloud detection method based on a time series of MODIS surface reflectance images', *International Journal of Digital Earth*, Vol. 6, sup1, pp.157–171.

Tricoli, U. (2015) *Electromagnetic scattering with the GDT-matrix method: an application to irregular ice particles in cirrus*, PhD Thesis, Ruperto-Carola-University of Heidelberg, Germany.

Wang, X., Parisien, M-A., Taylor, S.W., Candau, J-N., Stralberg, D., Marshall, G.A., Little, J.M. and Flannigan, M.D. (2017) 'Projected changes in daily fire spread across Canada over the next century', *Environment Research Letters*, Vol. 12, 025005, pp.1-12.

Wiggins, E.B., Czimczik, C.I., Santos, G.M., Chen, Y., Xu, X., Holden, S.R., Randerson, J.T., Harvey, C.F., Kai F.M. and Yud, L.E. (2018) 'Smoke radiocarbon measurements from Indonesian fires provide evidence for burning of millennia aged peat', *Proceedings for the National Academy of Sciences of the United States of America*, Vol. 115, No. 49, pp.12419–12424.

Wind, G., Platnick, S., King, M.D., Hubanks, P.A., Pavolonis, M.J., Heidinger, A.K., Yang, P. and Baum, B.A. (2010) 'Multilayer cloud detection with the MODIS near-Infrared water vapor absorption band', *Journal of Applied Meteorology and Climatology*, Vol. 58, No. 10, pp.2315–2333.

Xi, X., Natraj, V., Shia, R.L., Luo, M., Zhang, Q., Newman, S., Sander, S.P. and Yung, Y.L. (2015) 'Simulated retrievals for the remote sensing of $CO_2$, $CH_4$, CO, and $H_2O$ from geostationary orbit', *Atmospheric Measurement Technique*, Vol. 8, No. 11, pp.4817–4830.

Xiong, X., Wenny, B.N. and Barnes, W.L. (2009) 'Overview of NASA Earth Observing Systems Terra and Aqua moderate resolution imaging spectroradiometer instrument calibration algorithms and on-orbit performance', *Journal of Applied Remote Sensing*, Vol. 3, No. 1, p.032501.

Xiong, X., Wolfe, R., Barnes, W., Guenther, B., Vermote, E., Saleous, N. and Salomonson, V. (2010) 'Terra and Aqua MODIS design, radiometry, and geometry in support of land remote sensing', in *Land Remote Sensing and Global Environmental Change* (pp.133-164), Springer, New York.